\documentstyle[aps,prl,multicol,epsfig]{revtex}

\begin{document}
\title{Negative differential resistance in nanotube devices}
\author{Fran\c{c}ois L\'{e}onard and J. Tersoff}
\address{IBM Research Division, T. J. Watson Research Center\\
P.O. Box 218, Yorktown Heights, NY 10598}
\date{\today}
\maketitle

\begin{abstract}
Carbon nanotube junctions are predicted to exhibit negative differential
resistance, with very high peak-to-valley current ratios even at room
temperature. We treat both nanotube {\it p-n} junctions and undoped
metal-nanotube-metal junctions, calculating quantum transport through the
self-consistent potential within a tight-binding approximation. The undoped
junctions in particular may be suitable for device integration.
\end{abstract}


\begin{multicols}{2}
\narrowtext

The scaling of semiconductor devices to ever-smaller sizes is rapidly
approaching fundamental limits, spurring the exploration of new materials.
Carbon nanotubes (NTs) have particular appeal due to their small size and
unique mechanical and electronic properties; and some NT devices have
already been reported \cite{dekker}.

Recent theoretical work on NT devices has focused on simple operations such
as rectification \cite{leonard,odintsov,farajian}, based on thermal
excitation of carriers {\it over} a potential barrier. However, in NT devices
tunneling {\it through} the barrier
can actually dominate the transport \cite{leonard,leonard2}. Such tunneling
currents can lead to negative differential resistance (NDR), with a wide
range of potential device applications \cite{roy,ndr}.

Here we show that NT junctions are ideally suited to function as nanoscale
NDR device elements. We consider two very different devices --- a simple
{\it p-n} junction, and an entirely new device structure based on metal
contacts to an {\it undoped} NT.
The latter device relies on the nanoscale lateral size
of nanotubes and has no analog in bulk devices,
illustrating the exciting possibilities that nanotubes present.

In both cases, the direct gap and long tunneling length of the
NT contribute to a high peak current, while the strong carbon bonding and
small device size reduce the likelihood of any defect levels in the bandgap
contributing to excess valley current. Thus the predicted peak/valley
current ratios exceed by orders of magnitude those seen in existing devices.

Nanotube {\it p-n} junctions have been studied before \cite
{leonard,farajian,esfarjani}, although their potential as NDR devices has
not been recognized.
These simple devices provide an ideal testing ground for
general ideas about device operation. For technological applications,
however, one would prefer a device that does not require doping
and that can be integrated into a multi-level architecture.
The metal-NT-metal structure considered here
has precisely these desirable attributes.

We first consider a {\it p-n} junction made with a semiconducting
single-wall NT. Specifically, we treat a (17,0) zigzag NT with a radius of
0.66 nm. (See Ref.\ \cite{dekker} for notation and a general description of
the NT atomic structure.) Our qualitative results also apply to other
semiconducting NTs. We use a tight-binding Hamiltonian with one $\pi $
orbital per carbon atom and a nearest-neighbor matrix element of 2.5 eV \cite
{wildoer}, giving a direct band gap of 0.55 eV. We consider NT junctions
both in vacuum and embedded in a dielectric material ($\epsilon=$ 3.9).

Doping of NTs could be accomplished by insertion of atoms inside the tubes
\cite{miyamoto} or by substitution into the lattice \cite{subst},
and we have considered
both methods. We model dopants inside the tube by a line of charge on the
tube axis, either positive for {\it n} doping or negative for {\it p}.
(Replacing the line with discrete ions has negligible effect.) A reasonable
packing density corresponds to highly degenerate doping, with an atomic
doping fraction of about 0.01, and we use this value throughout our
calculations. Substitutional doping can be modeled (within a sort of
virtual-crystal approximation) by a uniform cylinder of charge on the tube.
I-V curves calculated with these two models agree within 1-2\% for all
applied voltages.

We calculate the current using the method of Ref.\ \cite{datta}. The NT is
divided into two semi-infinite ``leads'' and a ``scattering region'' 13.3 nm
in length. Within the scattering region we use the full self-consistent
potential $U(z)$, including applied bias and free-carrier screening. The
potentials in the leads are taken as constant, and equal to the potentials
at the boundaries of the scattering region;
and the scattering region is taken long enough to assure the accuracy
of this approximation.

To obtain $U(z)$, we self-consistently calculate the charge and potential
for a periodically repeated cell of 26.6 nm, consisting of {\it p} and {\it n%
} regions of equal size. The local density of states on each atomic site is
obtained by direct diagonalization of the Hamiltonian.
The charge on each site is given by integration of the product of the local
density of states and the Fermi function.

This standard method is only directly applicable in equilibrium, and must be
adapted for the presence of an applied voltage. In the limit of large
junction resistance (low current), the {\it p} and {\it n} regions are each
in internal equilibrium, but with Fermi levels that differ by the applied
voltage. We therefore calculate the charge using separate Fermi functions
for the two regions. (There is a region near the junction where the Fermi
level is undefined; but in the voltage range of interest, this region is
fully depleted and contributes negligible free charge regardless of which
Fermi level is used.) The accuracy of this approach is discussed further
below.

For a given charge $\sigma (z)$ (including both electronic and ionic
contributions), the electrostatic potential is $U(z)=\left( R/4\pi \epsilon
\right) \int \sigma \left( z^{\prime }\right) G(z-z^{\prime })dz^{\prime }$,
where $G(z-z^{\prime })$ is the electrostatic kernel for a cylinder \cite{G}
and $R$ is the NT radius.
[For a tube embedded in a dielectric ($\epsilon \ne 1$), the formula
neglects the presence of a hole in the dielectric.
A more accurate calculation in the context of the metal-NT-metal device is
presented below.]
In our numerical procedure, we start from a charge $\sigma (z)$ and obtain $%
U(z)$; the diagonal elements of the Hamiltonian are then shifted by $-eU(z)$%
, the electronic charge is re-calculated, and the procedure is iterated to
self-consistency.

The tight-binding formalism gives only the total charge associated with a
site. To calculate the potential, we must assume a particular spatial
distribution of the site charge.
In our calculation the charge associated with a ``ring" of sites is
distributed uniformly over a length $\xi$ of the NT cylinder. To test the
sensitivity of the results to the spatial distribution, we vary $\xi$ from
0.05 nm to 0.5 nm, which includes all physically reasonable values. Over this
large range the current varies less than 5\%. We expect that the effect of
varying the radial extent of the charge or lifting the approximation of
cylindrical symmetry would be similarly minor.

The Landauer-B\"uttiker formula \cite{datta} gives current
\begin{equation}
I=\frac{4e}{h}\int P(E)[F(E-eV/2) -F(E+eV/2)]dE ~.
\label{current}
\end{equation}
Here $P(E)$ is the electron transmission probability across the scattering
region at energy $E$, $V$ is the applied voltage,
and $F(E)$ the Fermi function. We keep
only the first valence and conduction bands in our calculation because
contributions from other bands are negligible here.

Figure 1(a) shows the electrostatic potential energy for a junction in
vacuum at zero bias.
The potential variation is quite large and sharp: the potential step is
1.08 eV (almost twice the band-gap, due to the highly degenerate
doping) and is largely localized to a region of less than 2 nm.

Figure 2 shows the calculated I-V curve for this device at room temperature.
For voltages from 0.25 to 0.6 V, the current decreases with increasing
voltage. Thus the device exhibits NDR, with an ``average'' value of $-55$ k$%
\Omega$ over this range. Moreover, the peak-to-valley current ratio is very
high, of order $10^4$.
The peak current is large because the NT has a direct gap and the valence
and conduction bands have the same rotational symmetry (analogous to having
the same transverse wavevector), allowing efficient direct tunneling. The
tunneling process does not require defects, phonons, or other scattering
mechanisms.

Above 0.6 eV, the only current is from thermionic
emission over the potential barrier. This gives an extremely low valley
current at room temperature. In conventional semiconductor junctions, there
is considerable additional current at the valley voltage due to
recombination via defect states in the bandgap \cite{sze}. For NTs we expect
a very low density of such defects, because of the very strong bonding
between carbon atoms. Moreover, there are only a relatively small number of
atoms in the actual device region, making it especially unlikely for a
defect to be present there.

Figure 2 also shows the I-V curve for the same device embedded in a
dielectric (modeled as discussed above). The qualitative behavior is
unchanged, with small shifts in the peak and valley voltages. The
peak-to-valley ratio is still $\sim$$10^4$.

Our calculation of the charge density is accurate at low current, so the I-V
curve has correct value and slope both at low voltage and at the valley
voltage. In the range of interest, the maximum error occurs near the current
peak. To quantify the accuracy, we consider a complementary approximation
that becomes accurate in the limit of high transmission through the junction
($P \rightarrow 1$), where a negligible fraction of the voltage drop occurs
across the junction. In this case the electrons moving left-to-right obey
the Fermi distribution of the left lead, while those moving right-to-left
obey the distribution of the right lead. Thus the total charge is well
approximated by occupying the states according to the arithmetic mean of the
two Fermi functions.

This approximation always gives a higher current in the range of interest,
because the potential step is affected very little by the voltage.
In particular, the peak current is increased by a factor of 2.
Since the actual transmission probability at the peak is $\lesssim 1/2$,
our calculations probably underestimate the peak current
but by considerably less than a factor of 2.
Thus the NDR performance is even better than that shown in Fig.~2.

The origin of the NDR
is similar to that in planar junctions\cite{sze}.
Under a small applied bias, as in Fig.~1(b), net current is generated due to
tunneling of electrons from filled conduction states on the {\it n} side to
empty valence states on the {\it p} side, with a high transmission
probability ($\sim$0.5). The current increases with the applied bias until
the Fermi levels align with the band edges on the opposite side of the
junction as shown in Fig.~1(c). (This condition for the maximum current is
different from that in bulk devices\cite{sze} because the NT density of
states peaks at the band edge.) Further increase in the voltage reduces the
range of energies where valence and conduction band states overlap, leading
to the NDR regime where current decreases with increasing voltage.
This NDR regime persists until the valence and conduction band edges on
opposite sides of the junction align, as in Fig.~1(d). At this point the
current across the device is at a minimum. For larger voltages, current
transport occurs only through thermal excitation of electrons over the
potential step (thermionic emission), and the device current increases
exponentially with increasing voltage.

We now consider an NDR device that does not require doping of NTs. As
illustrated in Fig.~3(a), the device consists of a semiconducting
single-wall NT with each end embedded in a different metal. The metals are
spaced apart by a layer of dielectric material.

Conceptually, one can imagine fabricating such a device beginning with a NT
growing vertically from a surface \cite{xu}, and sequentially
depositing a metal, a dielectric, and a second metal. The dielectric may be
replaced by a metal-oxidation step, and we assume that the materials do not
grow on or wet the NT.
Possible fabrication processes are discussed further below.

Because the NT does not form covalent bonds with the metal or dielectric, we
focus on the limit of weak metal-NT coupling. Then the matrix elements of
the NT Hamiltonian are unaffected by the presence of the metal, but charge
transfer between metal and NT (and the resulting electrostatic potential)
must still be included.

We take the metal on the right to have a workfunction larger than the NT
ionization potential, so that electrons are transferred from the NT valence
band to this metal. The metal on the left has a workfunction smaller than
the NT electron affinity, so that electrons are transferred from this metal
to the NT conduction band.
Deep within each contact, the population of the NT bands will be determined
by equilibrium with that metal, giving an ohmic contact.

The electrodes are modeled as semi-infinite ideal metals.
The embedded NT creates a cavity of radius $R+s$, where $s$ represents the
van der Waals separation between the NT and the metal. We assume a
separation of 0.3 nm, but varying $s$ between 0.2 and 0.5 nm has little
effect on our results.

For our numerical calculations we again treat a (17,0) NT, self-consistently
solving for the potential and charge on the tube at room temperature and
calculating the current, as for the {\it p-n} junction. The thickness of the
dielectric layer separating the electrodes is varied between 2 and 10 nm.
The two metal workfunctions are taken to be 1 eV above and below the NT
midgap (equivalent to the workfunction of a metallic NT).

The inset in Fig.~3(b) shows the self-consistent band alignment in
equilibrium. Because of charge transfer between the metal and NT,
the Fermi level is only about 0.1 eV above or below the NT band edge. As for
the case of the degenerately doped {\it p-n} junction, filled conduction
states overlap in energy with empty valence states, permitting tunneling and
leading to NDR.

The calculated I-V curves are shown in Fig.~3(b). The peak-to-valley ratio
for this device is as high as $10^7$ for the 2 nm device, and even the 10 nm
device has a peak-to-valley ratio of $10^5$. These are orders of magnitude
larger than for conventional planar devices \cite{sze}, and are comparable
or better than recently measured ratios in monolayers \cite{MW}.
Note that the device operation relies on controlling the potential via
the electrostatic boundary conditions, with changes of order 1 eV
over distances of order the tunneling length.
This is possible only because of the nanoscale lateral dimensions,
and has no analog in standard NDR devices.


Many strategies are possible for fabrication of the metal-NT-metal device;
and a sketch of one is shown in Fig.~4.
This is not intended as a realistic proposal for a specific process,
but merely to illustrate the issues and opportunities.
The process begins by defining parallel lines of
the first metal on a substrate, and laying the NT perpendicular to these
lines to create a suspended NT structure [Fig.~4(a)]. Then, more of the
first metal is deposited on the NT directly over the bottom contacts to
create embedded contacts [Fig.~4(b)]. (An alternative method is to suspend
the tube on insulating supports and grow the metal lines where the NT is
suspended.) To create the dielectric layer, either the surface of the first
metal is oxidized [Fig.~4(c)] or a dielectric is deposited as a blanket to
cover the whole system. Lines of the second metal are then deposited in
alignment with the first metal lines [Fig.~4(d)].

Fabrication of the devices described here poses many practical challenges,
but there appear to be no fundamental obstacles in principle.
The possibility of integrated fabrication,
together with the nanoscale device dimensions,
high peak-to-valley ratios, and other desirable properties of NTs,
makes these devices attractive candidates for nanoelectronics.

F.L.\ acknowledges support from the NSERC of Canada.

\begin{figure}[h]
\psfig{file=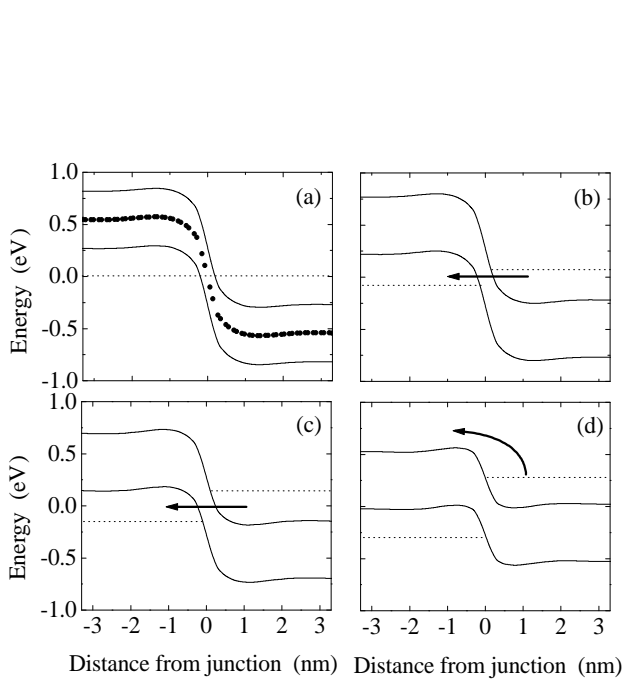,width=240pt,height=200pt}
\caption{Local valence and conduction band edges from the calculated
self-consistent potential for (a) $V=0$, (b) $V=0.1$, (c) $V=0.25$, and (d) $%
V=0.6$ Volts. Dotted lines are the Fermi levels.  Arrows indicate the
direction of electron flow. In panel (a), the electrostatic shifts of the
diagonal elements of the Hamiltonian are shown as dots. }
\end{figure}

\begin{figure}[h]
\psfig{file=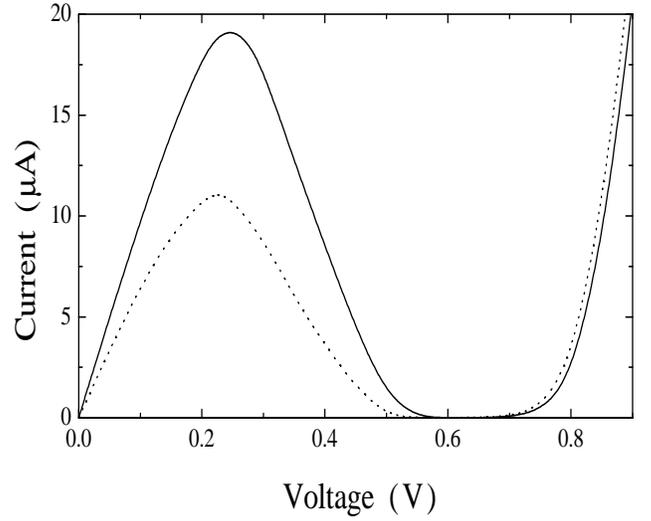,width=240pt,height=200pt}
\caption{I-V characteristics of the nanotube {\it p-n} junction. Solid line
is for a junction in vacuum, dotted is for a junction embedded in a
dielectric with $\protect\epsilon =3.9$.}
\end{figure}

\begin{figure}[h]
\psfig{file=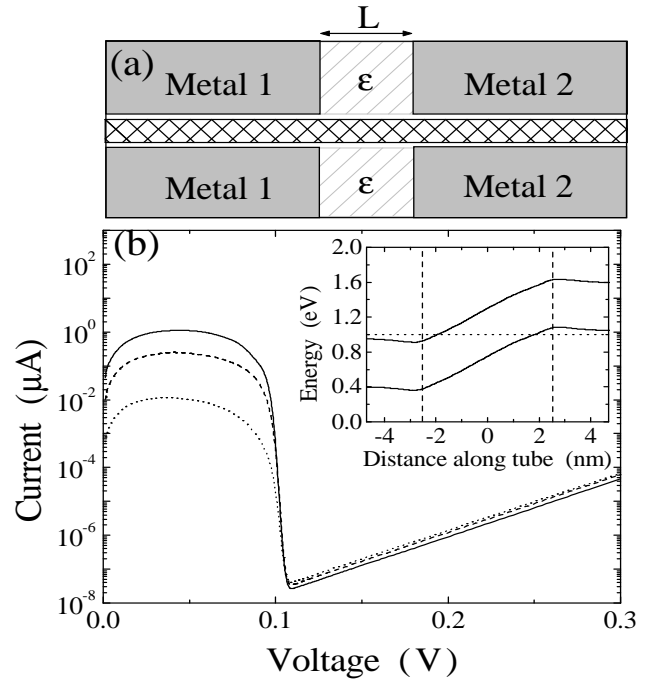,width=240pt,height=260pt}
\caption{(a) Cross-sectional view of metal-nanotube-metal device. (b)
Calculated I-V curves for this device. Solid, dashed, and dotted curves are
for electrode separations of $L=$ 2, 5, and 10 nm respectively.
The inset in (b) shows the self-consistent band diagram for
the $L=$ 5 nm case in equilibrium ($V=0 $).}
\end{figure}

\begin{figure}[h]
\psfig{file=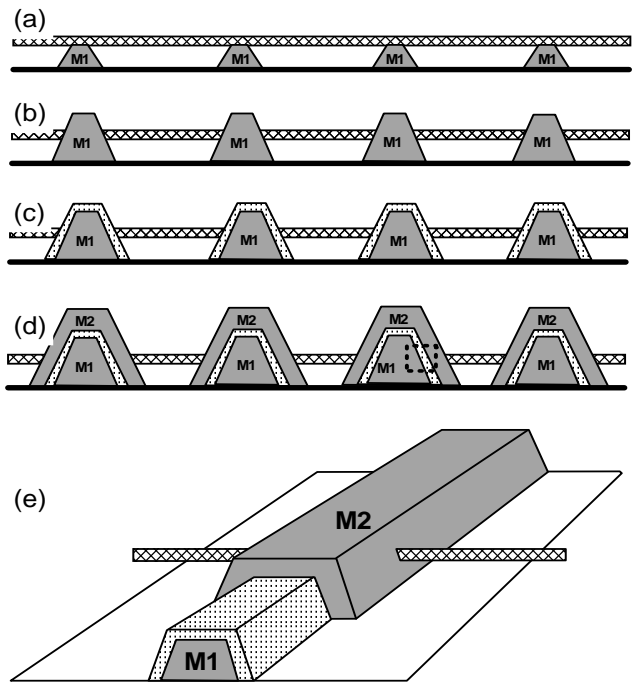,width=240pt,height=260pt}
\caption{Sketch of a possible fabrication sequence (a-d) for the
metal-NT-metal device, in cross-sectional view. (e) Perspective view of one
of the finished devices. The dotted box in (d) shows the device region (as
in Fig.\ 3a).}
\end{figure}

\end{multicols}

\end{document}